\documentclass[conference]{IEEEtran}

\IEEEoverridecommandlockouts

\usepackage{cite}
\usepackage{amsmath,amssymb,amsfonts}
\usepackage{graphicx}
\usepackage{textcomp}
\usepackage{xcolor}
\usepackage{url}
\usepackage{algorithm}
\usepackage{algorithmic}
\usepackage{float}
\usepackage{placeins}
\usepackage{stfloats}

\def\BibTeX{{\rm B\kern-.05em{\sc i\kern-.025em b}\kern-.08em
    T\kern-.1667em\lower.7ex\hbox{E}\kern-.125emX}}

\begin{document}

\title{Real-Time 2D LiDAR Object Detection Using Three-Frame RGB Scan Encoding}

\author{
\IEEEauthorblockN{Soheil Behnam Roudsari$^{1}$,
Alexandre S. Brand{\~a}o$^{2}$,
and Felipe N. Martins$^{1}$\thanks{Supplementary material (code, trained weights, and demo video) is available at: \protect\url{https://github.com/soheilbh/2d-lidar-identification} and \protect\url{https://www.youtube.com/watch?v=sjO1z04g8Jg} This work has been submitted to the IEEE for possible publication. Copyright may be transferred without notice, after which this version may no longer be accessible.}}
\IEEEauthorblockA{$^{1}$Sensors and Smart Systems Group, Institute of Engineering \\
Hanze University of Applied Sciences, Groningen, The Netherlands \\
s.behnam.roudsari@st.hanze.nl, fe.nascimento.martins@pl.hanze.nl}
\IEEEauthorblockA{$^{2}$N\'ucleo de Especializa\c{c}\~ao em Rob\'otica, Departamento de Engenharia El\'etrica \\
Universidade Federal de Vi\c{c}osa, Vi\c{c}osa, MG 36570-900, Brazil \\
alexandre.brandao@ufv.br}
}

\maketitle

\begin{abstract}
  Indoor service robots need perception that is robust, more privacy-friendly than RGB video, and feasible on embedded hardware. We present a camera-free 2D LiDAR object detection pipeline that encodes short-term temporal context by stacking three consecutive scans as RGB channels, yielding a compact YOLOv8n input without occupancy-grid construction while preserving angular structure and motion cues. Evaluated in Webots across 160 randomized indoor scenarios with strict scenario-level holdout, the method achieves 98.4\% mAP@0.5 (0.778 mAP@0.5:0.95) with 94.9\% precision and 94.7\% recall on four object classes. On a Raspberry Pi~5, it runs in real time with a mean post-warm-up end-to-end latency of 47.8~ms per frame, including scan encoding and postprocessing. Relative to a closely related occupancy-grid LiDAR--YOLO pipeline reported on the same platform, the proposed representation is associated with substantially lower reported end-to-end latency. Although results are simulation-based, they suggest that lightweight temporal encoding can enable accurate and real-time LiDAR-only detection for embedded indoor robotics without capturing RGB appearance.
\end{abstract}

\begin{IEEEkeywords}
2D LiDAR, object detection, YOLOv8, embedded robotics, temporal encoding, privacy-friendly sensing
\end{IEEEkeywords}

\section{Introduction}

Indoor service robots must detect and react to objects reliably while running on small onboard computers. Modern camera-based detectors achieve strong accuracy \cite{wang2024yolov9}, but vision can be sensitive to lighting and it raises privacy concerns in human-populated indoor spaces \cite{lintvedt2023thermal}. In many real deployments (homes, hospitals, offices), perception must also be more privacy-friendly than continuous RGB video. These constraints motivate sensing pipelines that reduce reliance on RGB appearance while remaining efficient enough for embedded CPUs.

LiDAR provides geometric structure without capturing visual appearance, making it a practical privacy-friendly option for indoor robots. High-end 3D multi-beam LiDAR can support rich perception, but it typically increases hardware cost, size, power consumption, and integration complexity, which is often unsuitable for low-cost service platforms. In contrast, 2D LiDAR is widely available and inexpensive, but its sparse measurements make multi-class object detection more challenging than camera-based detection.

Recent learning-based methods indicate that 2D LiDAR can support semantic perception \cite{najem2024object}. However, a key gap remains: achieving accurate multi-class detection with \emph{low end-to-end latency} on embedded hardware. In particular, pipelines that convert scans into dense occupancy grids increase preprocessing overhead and input size, which can limit real-time performance.

This paper presents a camera-free 2D LiDAR object detection pipeline that encodes short-term temporal context using a compact multi-frame representation. We rasterize three consecutive 2D scans into binary images and stack them as RGB channels to form a tensor compatible with YOLOv8n. This preserves the scan's angular structure and provides short-term motion cues, while avoiding occupancy-grid construction.

The main contributions are:
\begin{enumerate}
    \item A compact multi-frame 2D LiDAR encoding that stacks three consecutive scans as RGB channels, injecting short-term temporal cues without pose alignment;
    \item A YOLOv8n-based LiDAR-only detector evaluated under strict scenario-level holdout in simulation;
    \item An end-to-end embedded implementation demonstrating real-time performance on low-cost hardware;
    \item A comparison to an occupancy-grid LiDAR--YOLO baseline, highlighting latency and representation-cost implications of the proposed encoding.
\end{enumerate}

The remainder of this paper is organized as follows: Section~II reviews related work; Section~III describes the proposed method; Section~IV presents experiments; and Section~V concludes.

\section{Related Work}

Early 2D LiDAR perception for mobile robots often relied on geometric preprocessing such as clustering \cite{seckin2021pedestrian} and hand-crafted rules \cite{fagundes2024analytical}. These approaches are efficient and interpretable, but they require careful tuning and do not scale well to diverse environments and richer object categories.

Learning-based methods have shown that 2D LiDAR can support semantic perception when the ordered angular structure of scans is preserved. Temporal context is also known to improve robustness in sparse range data. For example, DROW and DR-SPAAM fuse multiple consecutive scans to stabilize detection under occlusions and partial views without requiring expensive scan matching \cite{beyer2018deepperson,jia2020drspaam}. However, these systems typically focus on a narrow set of targets (often people) and do not directly address YOLO-style multi-class object detection with bounding boxes.

A strategy for applying CNN detectors to 2D LiDAR is to convert scans into dense occupancy-grid images. Najem et al. (2025) follow this approach by rasterizing scans and applying YOLOv8n, but they report high end-to-end latency on Raspberry Pi~5, illustrating the preprocessing and input-size costs of grid construction \cite{najem2024object}. In contrast, other work processes scans closer to their native representation using lightweight models, such as 2DLaserNet \cite{kaleci20212dlasernet} and TinyLidarNet \cite{zarrar2024tinylidarnet}, demonstrating that compact architectures can run on constrained hardware. These methods, however, are not designed as multi-class 2D object detectors in an image-style (YOLO) formulation.

In parallel, 3D LiDAR detection commonly uses BEV-style encodings and achieves strong results, but it often assumes expensive multi-beam sensors and higher compute budgets that are often incompatible with low-cost indoor service robots. Finally, privacy constraints in indoor environments motivate sensing that avoids RGB appearance capture; in this context, 2D LiDAR provides geometry-only measurements and is commonly treated as more privacy-friendly than video.

Our work builds on these directions by avoiding occupancy-grid construction and introducing a minimal temporal encoding (three scans stacked as RGB channels) that remains compatible with fast CNN detectors while keeping the input compact.

\section{Methods}
This section describes the proposed 2D LiDAR-only object detection pipeline, including the input representation, temporal encoding, dataset generation, training configuration, and deployment.

\subsection{System Overview}
We propose an efficient object detection pipeline based entirely on 2D LiDAR, designed for real-time execution on embedded CPUs. Compared with prior LiDAR--YOLO pipelines that first construct occupancy grids \cite{najem2024object}, our approach encodes scans directly into a compact RGB-style tensor that can be processed by standard CNN detectors.

The core idea is to stack three consecutive 2D scans as the three channels of a small raster image. This preserves the ordered angular structure of the scan while injecting short-term temporal cues (motion and viewpoint change) without requiring odometry alignment. The resulting input is compact ($64\times384\times3$) and avoids the preprocessing cost and large input sizes typical of occupancy-grid pipelines (often resized to $640\times640$).

Algorithm~\ref{alg:runtime_pipeline} summarizes the end-to-end runtime pipeline.
\begin{algorithm}[!t]
\caption{Online 2D LiDAR Inference Pipeline}
\label{alg:runtime_pipeline}
\begin{algorithmic}[1]
\STATE \textbf{Input:} raw 2D LiDAR scan at time $t$
\STATE \textbf{State:} FIFO buffer $\mathcal{B}$ holding the last 3 rasters
\STATE \textbf{Output:} bounding boxes and class scores at time $t$
\STATE Rasterize scan and pad to $64\times384$
\STATE Update $\mathcal{B}$ with the newest raster (keep 3)
\STATE Stack $\mathcal{B}$ into an RGB tensor ($64\times384\times3$)
\STATE Run YOLOv8n inference on the RGB tensor
\STATE Return detections (boxes and class scores)
\end{algorithmic}
\end{algorithm}

\subsection{Temporal Data Encoding}
A 2D LiDAR scan is a sequence of angle--range measurements
$\{(\theta_i,\rho_i)\}_{i=1}^{N}$, where $\theta_i$ is the bearing and $\rho_i$ is the measured distance.

\textbf{Rasterization.}
Each scan is converted into a sparse binary raster $\mathbf{I}_t \in \{0,255\}^{64\times360}$, where columns represent angle bins and rows represent discretized range bins. We use 360 angular bins (1 degree per bin). The angular bin index is computed as
\begin{equation}
a_b \;=\; \mathrm{mod}\!\left(\mathrm{round}(\theta\,[^\circ]),\,360\right),
\end{equation}
i.e., angles are rounded to the nearest degree and wrapped modulo 360.

For the range axis, we discretize into 64 bins with a fixed full-scale value $R_{\max}=4.0$~m. The simulator sensor range is 0.12--3.5~m; measurements that are infinite or out-of-range are replaced with the sentinel $R_{\max}$ (this sentinel is used only to handle invalid returns consistently, not to claim measurements beyond the nominal range). The range-bin index is computed as:
\begin{equation}
r_b \;=\; \left\lfloor \mathrm{clip}\!\left(\rho \cdot \frac{63}{R_{\max}},\,0,\,63\right) \right\rfloor,
\end{equation}
where $\rho$ is in meters. For each measurement, the pixel at $(r_b,a_b)$ is set to 255, and all other pixels remain 0.

\textbf{Axis convention.}
In the raster, the horizontal axis (columns) corresponds to angle bins $a_b\in[0,359]$, and the vertical axis (rows) corresponds to range bins $r_b\in[0,63]$. We use this convention consistently for label projection and bounding boxes.

\textbf{Padding to match CNN stride.}
YOLO backbones commonly assume spatial dimensions divisible by 32. To avoid rescaling the scan raster (which would distort the angular structure), we pad $\mathbf{I}_t$ with 24 zero columns to obtain $\mathbf{I}_t \in \{0,255\}^{64\times384}$.

\textbf{Three-frame temporal stacking.}
To encode short-term temporal context, we maintain a first-in, first-out (FIFO) buffer of three consecutive rasters at times $t-2$, $t-1$, and $t$. These are stacked into an RGB tensor
$\mathbf{I}^{\text{RGB}}_t \in \{0,255\}^{64\times384\times3}$:
\begin{equation}
\mathbf{I}^{\text{RGB}}_t(r,a,:) \;=\; \big[\mathbf{I}_{t-2}(r,a),\;\mathbf{I}_{t-1}(r,a),\;\mathbf{I}_{t}(r,a)\big].
\end{equation}
This encoding injects motion cues and short-term viewpoint variation without scan matching or pose alignment, keeping preprocessing lightweight.

\begin{figure}[!t]
    \centering
    \includegraphics[width=0.48\textwidth]{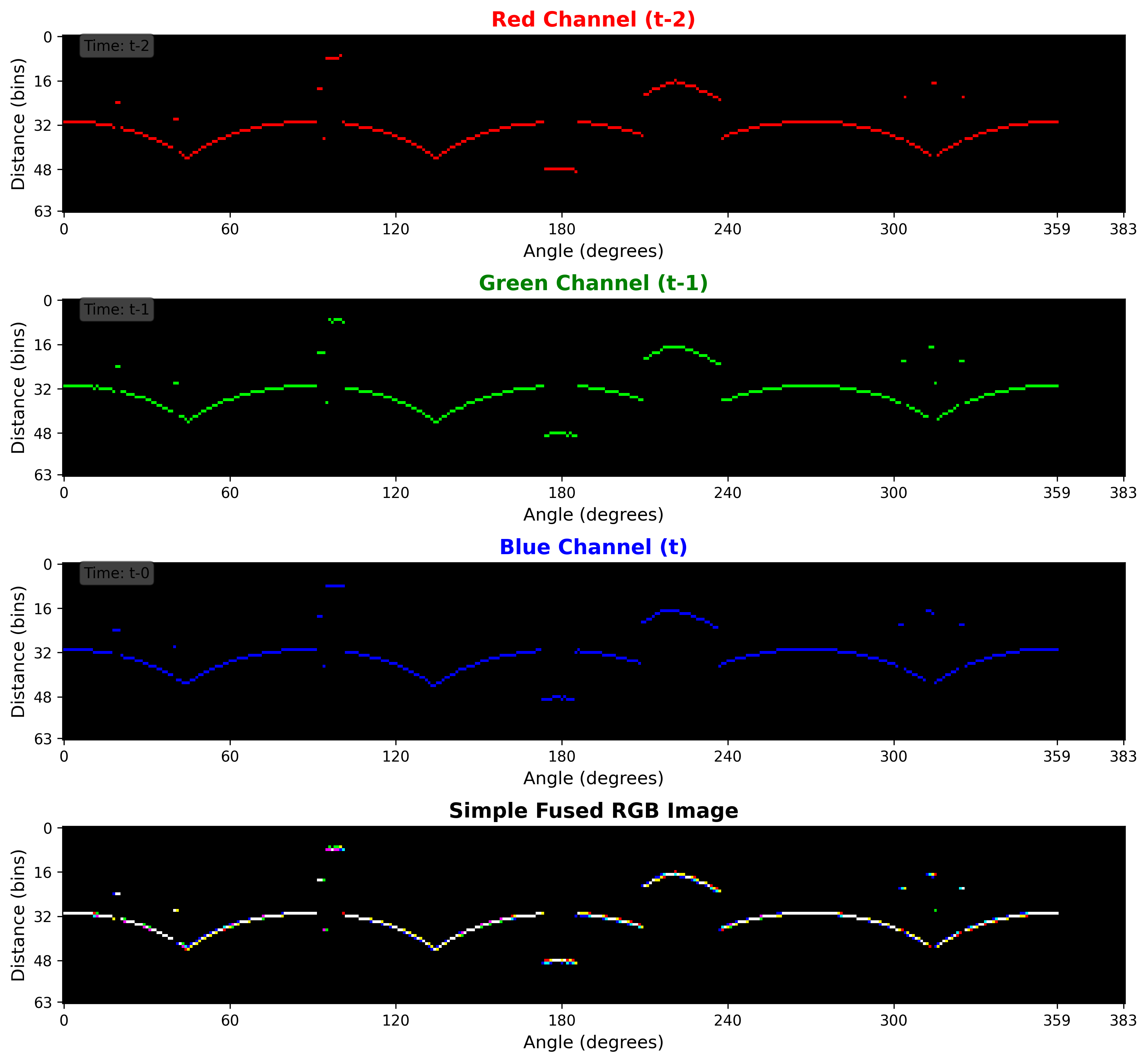}
    \caption{RGB-encoded LiDAR input: three consecutive scans are rasterized and stacked (red: $t{-}2$, green: $t{-}1$, blue: $t$) to form a compact tensor used by YOLOv8n.}
    \label{fig:rgb_encoding}
\end{figure}

Figure~\ref{fig:rgb_encoding} visualizes the per-channel stacking and a simple fused view.

\subsection{Alternative Encoding Exploration}
\begin{figure}[!ht]
  \centering
  \includegraphics[width=0.48\textwidth]{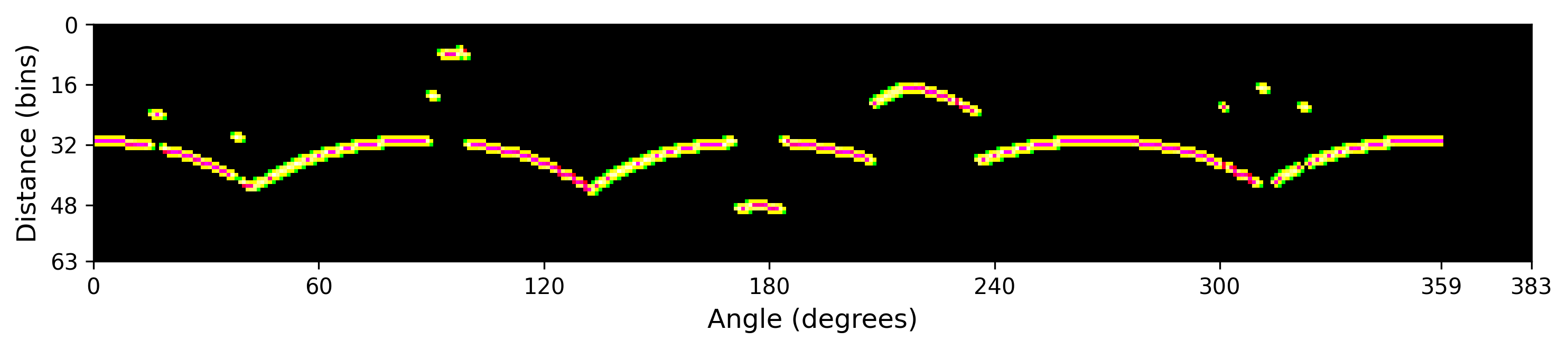}
  \caption{Visualization of an explored ``aligned fused'' encoding using a 5-frame buffer (hit counts, edges, and density). This was not used in the final pipeline due to added cost without clear accuracy gains.}
  \label{fig:alignedfused}
\end{figure}

During design iterations, we evaluated a more complex encoding that attempted to compensate for robot rotation between scans. This variant used a 5-frame buffer, applied yaw-based column shifts, and encoded (i) hit counts, (ii) Sobel edges, and (iii) point density into the three channels (illustrated in Fig.~\ref{fig:alignedfused}).

Although this produced visually cleaner temporal fusion, it increased preprocessing cost and memory usage. In an ablation experiment (same train/val split and YOLOv8n settings), the aligned 5-frame encoding did not improve mAP and increased preprocessing time; we therefore use the unaligned three-frame stacking in the final system to maximize end-to-end efficiency.

\subsection{Automated Dataset Generation}
\begin{figure}[!t]
    \centering
    \includegraphics[width=0.48\textwidth]{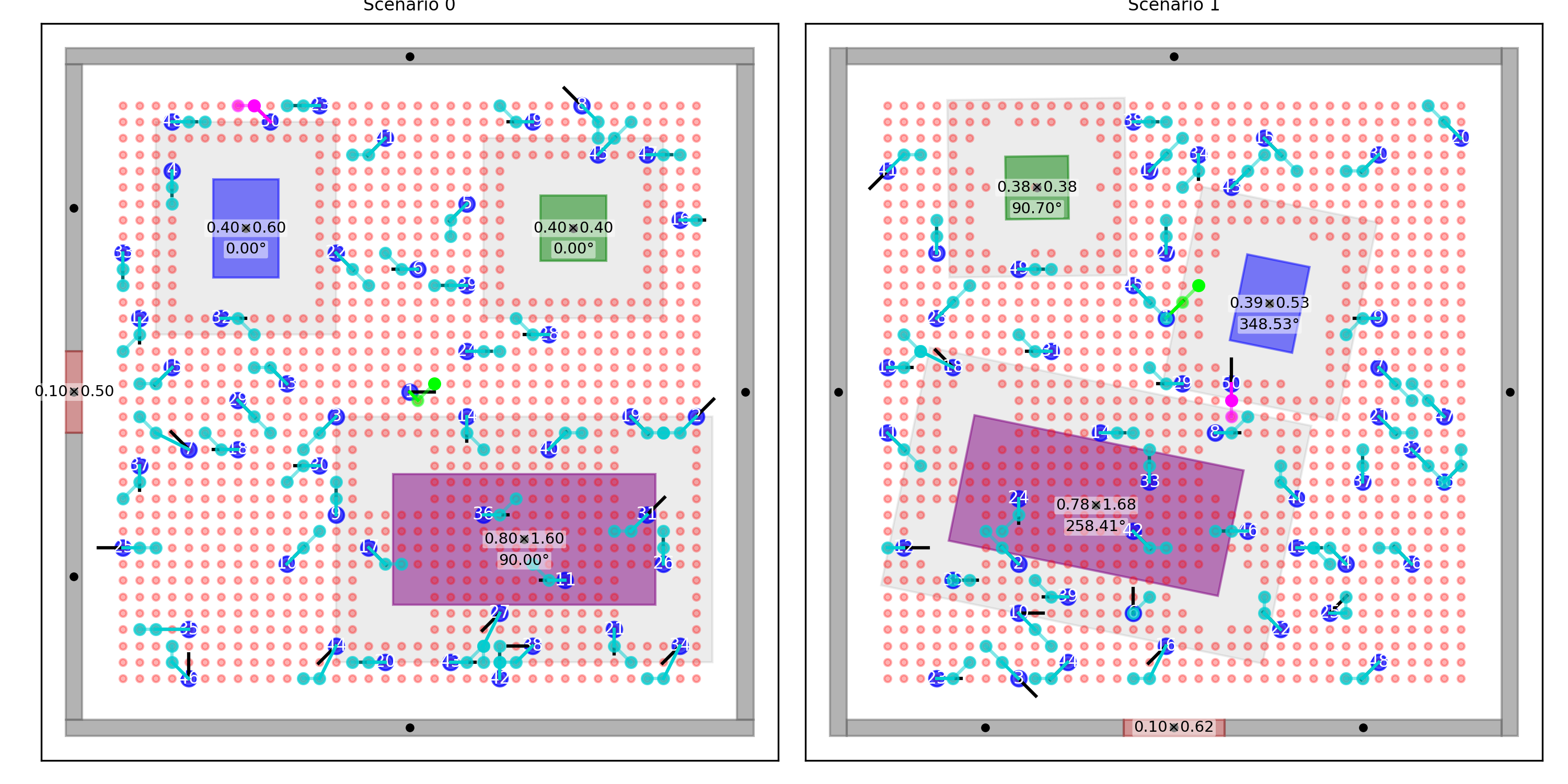}
    \caption{Example randomized scenarios with 90 predefined robot positions. Object types are shown as colored rectangles (box: blue, chair: green, desk: purple, doorframe: red) and robot waypoints as blue dots.}
    \label{fig:scenario_positions}
\end{figure}

\begin{figure}[!t]
    \centering
    \includegraphics[width=0.45\textwidth]{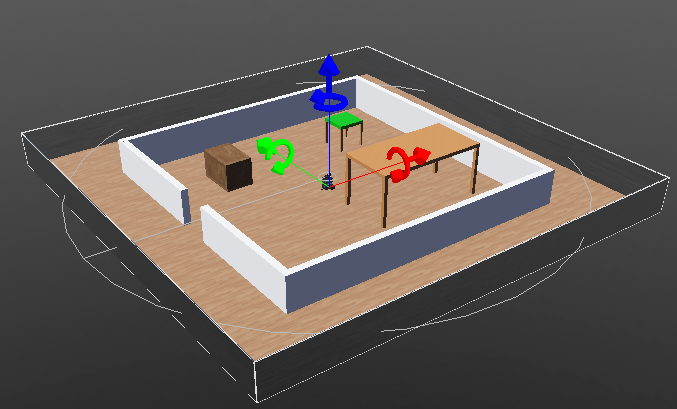}
    \caption{3D view of the Webots simulation environment showing a typical indoor scenario with objects (box, table, chair, doorframe) and the room structure.}
    \label{fig:simulation_3d}
\end{figure}

We generated training data in simulation using Webots R2023b. Each data collection run consists of $N$ randomized scenarios and $M$ predefined robot positions per scenario, producing $N\times M$ sampling episodes. At each position, the robot executes a short motion and we record multiple consecutive scans, yielding multiple labeled frames per episode. Fig.~\ref{fig:scenario_positions} shows example simulation scenarios, each containing predefined robot positions and randomized object configurations (chairs, desks, boxes, doorframes). Fig.~\ref{fig:simulation_3d} provides a 3D view of the simulation environment.

To improve diversity, the \emph{movable} objects (chair, desk, box) are randomized in planar translation and yaw, and their planar footprint is scaled by $\pm20\%$ (scale factors sampled uniformly in $[0.8,1.2]$). In Webots, the chair and desk classes are instantiated using the built-in \texttt{Table} model (rectangular tabletop supported by four legs); we use different nominal footprints (chair: $0.4\times0.4$~m, desk: $0.8\times1.6$~m) and leg thickness settings (\texttt{feetSize} $0.03$~m vs. $0.05$~m), with isotropic scaling for chairs and independent width/height scaling for desks. The box class uses the \texttt{WoodenBox} cuboid model with nominal footprint $0.4\times0.6$~m and independent width/height scaling. Doorframes correspond to the doorway opening in the enclosing walls; across scenarios, the room layout is rotated by multiples of $90^\circ$, so the doorway can appear on any side. We label doorframes using a thin rectangular proxy centered at the doorway with fixed thickness $0.1$~m and height sampled uniformly in $[0.5,1.0]$~m. Each training input is formed by stacking three consecutive scans as described above.

\begin{figure}[!t]
    \centering
    \includegraphics[width=0.48\textwidth]{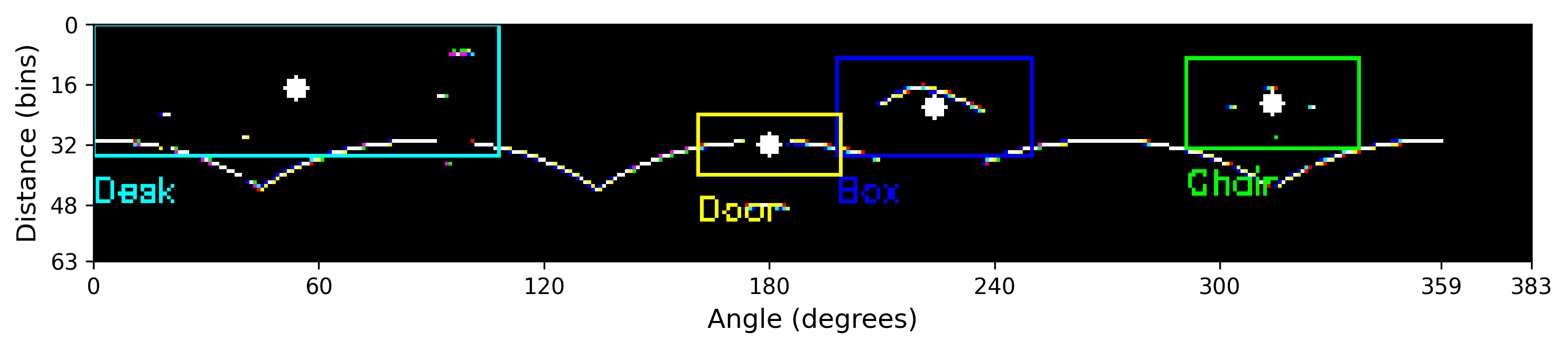}
    \caption{Example RGB-fused LiDAR input with automatically generated bounding boxes for the four object classes.}
    \label{fig:rgb_labels}
\end{figure}

\textbf{Label generation.}
Ground-truth labels correspond to time $t-1$ (the middle scan in the 3-frame buffer). Using simulator metadata and the robot pose at $t-1$, object corners are transformed into the robot frame and projected into raster coordinates (range-bin, angle-bin). Axis-aligned YOLO-format bounding boxes $(x_c,y_c,w,h)$ are then produced in normalized raster coordinates. Boxes are clipped to the angular bounds $[0,359]$; objects crossing the $0^\circ/359^\circ$ discontinuity are therefore represented by a clipped box rather than being split into two boxes. Fig.~\ref{fig:rgb_labels} shows an example RGB-fused input with automatically generated bounding boxes for all four object classes.

Across all episodes, this process produced 768{,}897 encoded samples from 160 scenarios evaluated at 90 robot positions each.

\subsection{Model Training}
We trained a YOLOv8n detector using the Ultralytics implementation (v8.3.152) in PyTorch. YOLOv8n is the nano variant of YOLOv8, chosen for embedded feasibility (approximately 3M parameters). The input resolution is fixed at $64\times384$, and rectangular training (rect=True) is enabled to preserve the native LiDAR aspect ratio.

Because LiDAR rasters have constrained geometry, we disable strong image augmentations (mosaic, mixup, cutmix, copy-paste). We apply only lightweight flips: a left--right flip along the angular axis (50\% probability) and an up--down flip along the range axis (20\% probability). The range-axis flip is not physically realizable for a fixed 2D sensor; we treat it strictly as an image-space regularizer and discuss it as a sim-to-real risk in Section~IV-D.

Training was performed on a server with an NVIDIA H100 GPU (80~GB VRAM). The model was trained for 120 epochs with batch size 1{,}024 (enabled by GPU memory and dataset caching). Optimization uses AdamW with an initial learning rate of 0.001 and cosine learning-rate scheduling, with early stopping based on validation mAP.

To evaluate generalization, we split the dataset at the \emph{scenario level}: 85\% of scenarios for training (629{,}591 samples), 10\% for validation (72{,}528 samples), and 5\% for testing (38{,}321 samples). This ensures that spatial layouts and object configurations in the test set are unseen during training.

\subsection{Deployment}
The trained model (.pt) was deployed on multiple platforms to validate embedded feasibility. On Raspberry Pi~5 and MacBook Air M2, inference runs using the Ultralytics runtime. For Raspberry Pi~3, the model is exported to TorchScript to enable inference on limited resources (functional but not real-time).

At deployment, the model consumes the $64\times384\times3$ RGB LiDAR tensor directly and outputs class scores and bounding boxes in a single forward pass, requiring only the lightweight encoding step described above. This tensor is formed by mapping each raw polar return $(\theta,\rho)$ into a 2D raster index $(r_b,a_b)$ via angular and range quantization (Eqs.~(1)--(2)), yielding an image plane where columns are angle bins and rows are range bins.

\section{Experimental Results}
This section evaluates the proposed method in simulation, reports detection accuracy and embedded runtime, and contextualizes the representation cost relative to prior work.

\subsection{Experimental Setup}
All evaluation data was generated in Webots R2023b on a MacBook Air M2. We created 160 unique randomized indoor scenarios and evaluated each from 90 predefined robot positions, yielding 14,400 sampling episodes. Labels are generated automatically from simulator metadata (object poses and dimensions) and robot pose as described in Section~III-D; no manual annotation is used. These episodes produced 768,897 labeled RGB-tensor inputs through the procedure described in Section~III-D. End-to-end simulation and labeling required approximately 18~hours.

The simulated 2D LiDAR sensor (LDS-01) provides 360$^\circ$ coverage at 10~Hz. Scans are assigned to 360 angular bins for rasterization. The effective measurement range is 0.12--3.5~m; infinite or out-of-range returns are replaced with 4.0~m and mapped to the far range bin.

We evaluate runtime on three platforms representing different compute budgets:
\begin{itemize}
    \item \textbf{Raspberry Pi~5 (8~GB RAM):} Broadcom BCM2712 quad-core Cortex-A76 @ 2.4~GHz;
    \item \textbf{MacBook Air M2 (8~GB unified memory):} Apple M2 (8-core CPU, 10-core GPU);
    \item \textbf{Raspberry Pi~3 Model B+ (1~GB RAM):} Broadcom BCM2837B0 quad-core Cortex-A53 @ 1.4~GHz (TorchScript).
\end{itemize}

\subsection{Detection Performance}
On the held-out test set (strict scenario-level holdout), the model achieves 0.984 mAP@0.5 and 0.778 mAP@0.5:0.95, indicating reliable detection and classification in unseen scenario layouts.

\begin{figure}[htbp]
  \centering
  \includegraphics[width=0.35\textwidth]{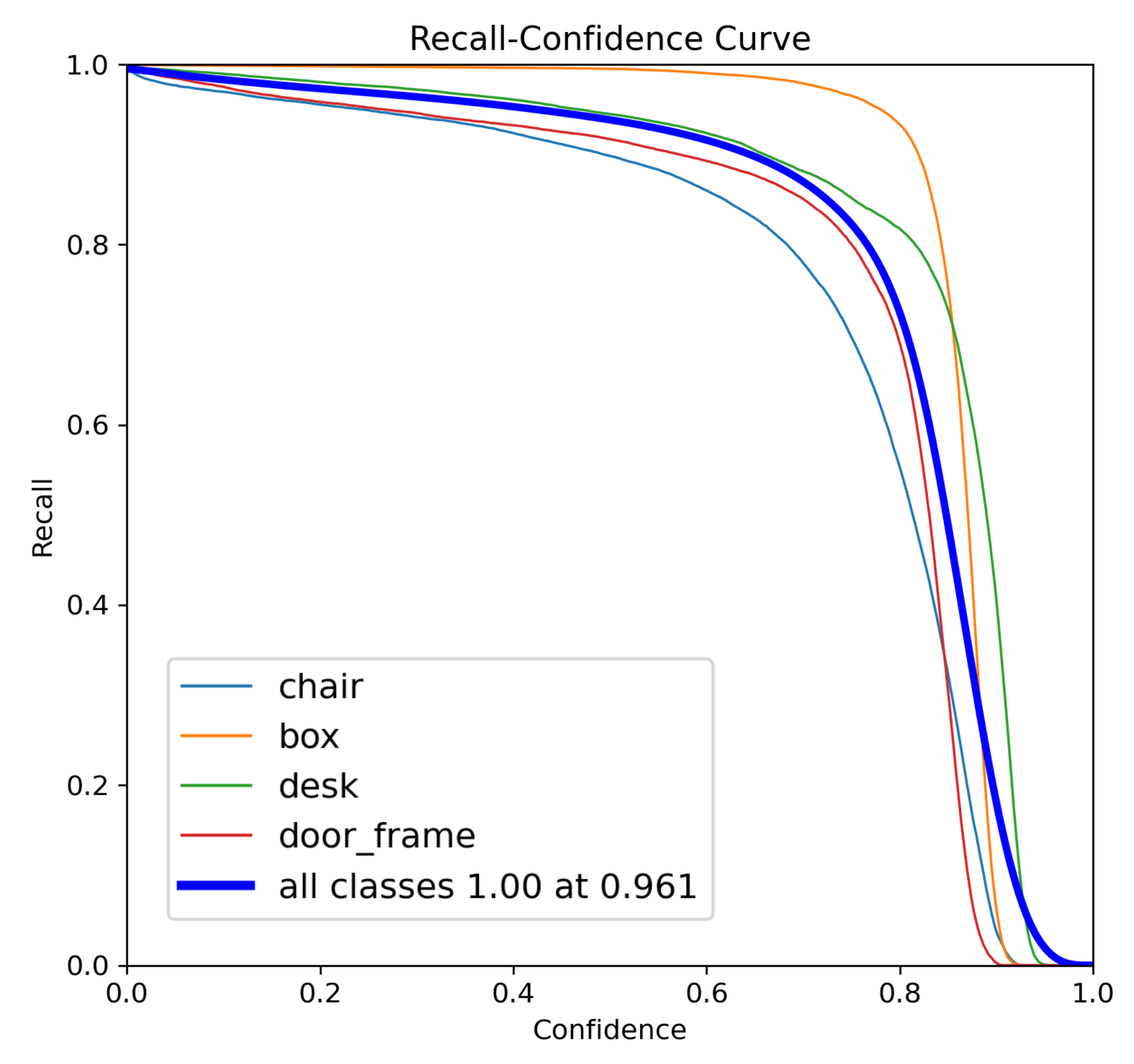}\\[0.5ex]
  \includegraphics[width=0.35\textwidth]{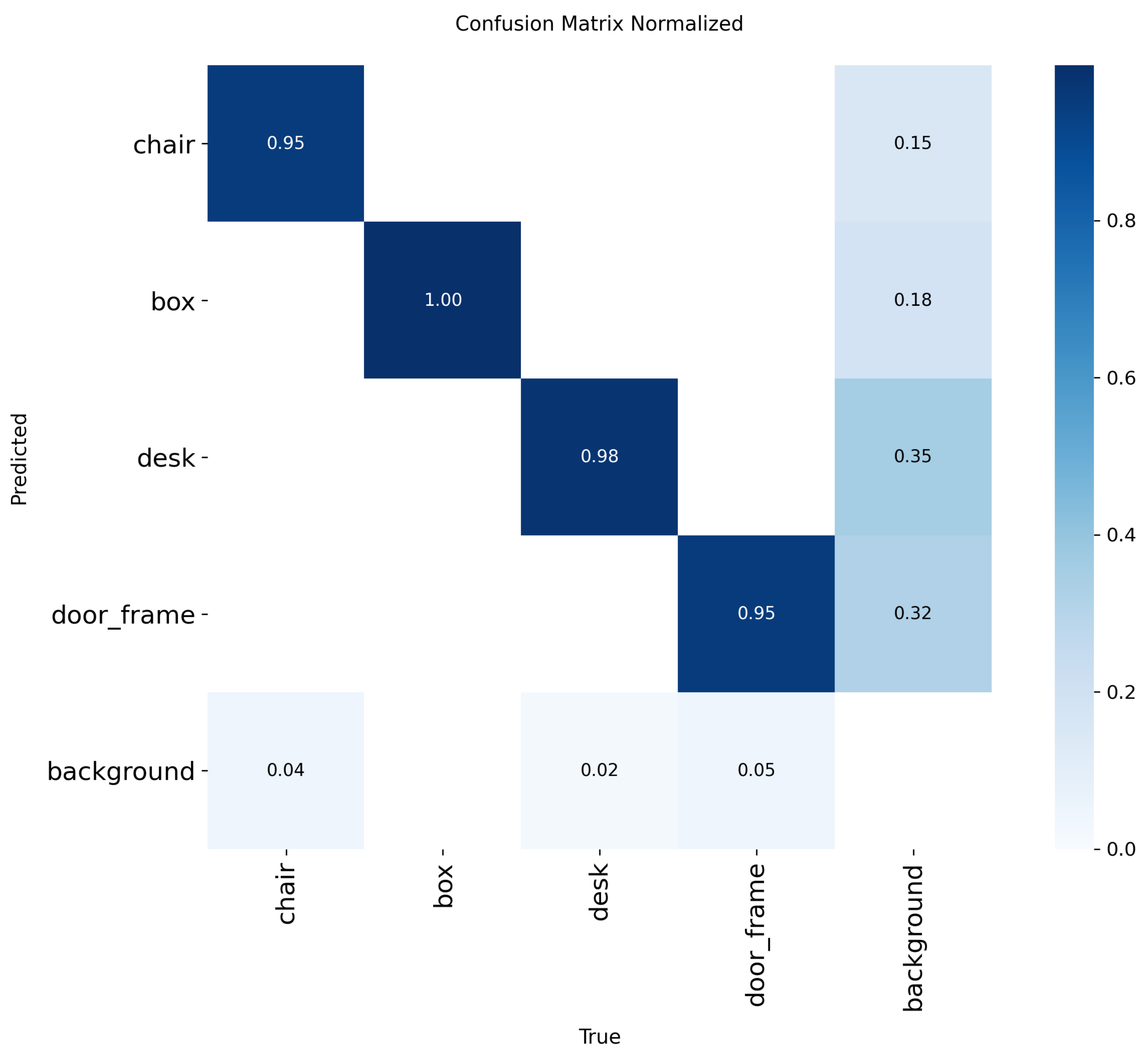}
  \caption{Top: Recall--Confidence curve. Bottom: Normalized confusion matrix.}
  \label{fig:perf_images}
\end{figure}

Table~\ref{tab:perclass} reports per-class precision, recall, and mAP. Performance is consistently high across all four classes. Most errors occur between \emph{desk} and \emph{doorframe}, which can produce similar partial geometric profiles in 2D LiDAR from certain viewpoints. Fig.~\ref{fig:perf_images} shows the recall--confidence curve and normalized confusion matrix, which further illustrate these performance characteristics. Background false detections remain very low ($<1\%$), indicating strong class separation.

\begin{table}[htbp]
\caption{Per-Class Detection Metrics on Test Set}
\begin{center}
\begin{tabular}{lcccc}
\hline
\textbf{Class} & \textbf{Precision} & \textbf{Recall} & \textbf{mAP@0.5} & \textbf{mAP@0.5:0.95} \\
\hline
Chair & 0.986 & 0.913 & 0.981 & 0.746 \\
Box & 0.957 & 0.996 & 0.993 & 0.780 \\
Desk & 0.922 & 0.954 & 0.982 & 0.822 \\
Doorframe & 0.930 & 0.926 & 0.980 & 0.763 \\
\hline
\textbf{All (mean)} & \textbf{0.949} & \textbf{0.947} & \textbf{0.984} & \textbf{0.778} \\
\hline
\end{tabular}
\label{tab:perclass}
\end{center}
\end{table}

\subsection{Inference Time Analysis}
We measure end-to-end CPU latency with batch size 1 and FP32 inference, including scan encoding, the YOLOv8n forward pass, and non-maximum suppression (NMS). For each platform, we perform one warm-up inference and report the mean latency over 100 consecutive frames. We target $<50$~ms end-to-end latency to support $\approx20$~Hz perception updates, which supports typical indoor navigation and control loops.

Table~\ref{tab:inference} summarizes the measured runtimes. On Raspberry Pi~5, the pipeline achieves a mean end-to-end latency of 47.8~ms per frame, meeting the real-time target while leaving computational headroom for other onboard modules (e.g., localization and planning). The MacBook Air M2 achieves substantially lower latency, whereas the Raspberry Pi~3 (TorchScript) remains functional but far from real time.

\begin{table}[htbp]
\caption{Inference Time Across Hardware Platforms}
\begin{center}
\begin{tabular}{lcc}
\hline
\textbf{Platform} & \textbf{Mean Time (ms)} & \textbf{Derived FPS} \\
\hline
MacBook Air M2 & 6.2 & 161.29 \\
Raspberry Pi~5 & 47.8 & 20.92 \\
Raspberry Pi~3 (TorchScript) & $\sim$2000 & $\sim$0.50 \\
\hline
\end{tabular}
\label{tab:inference}
\end{center}
\noindent \footnotesize Derived FPS values are computed as \(1/\text{mean latency}\). Timings include scan encoding, inference, and NMS, and are reported as the mean over 100 consecutive frames after one warm-up inference.
\end{table}

\subsection{Limitations and Threats to Validity}
This study is evaluated in simulation; real 2D LiDAR sensors introduce noise, missing returns, and calibration effects that may reduce performance under domain shift. Labels are derived from simulator metadata, which removes annotation ambiguity and may inflate reported metrics relative to manually labeled real-world datasets. The task considers four mostly static indoor object classes and does not evaluate dynamic agents (e.g., humans). Temporal stacking is performed without pose alignment and may degrade under fast robot rotation or highly dynamic scenes. Finally, while 2D LiDAR avoids RGB appearance capture, geometric signals may still enable \emph{behavioral} inference in some deployments; we therefore describe the approach as privacy-friendly rather than privacy-preserving.

\subsection{Comparison with Prior Work}
Table~\ref{tab:comparison} summarizes reported results from Najem et al.~\cite{najem2024object} alongside our method. Because Najem et al. evaluate on real data with manual labels while our results are simulation-based with automatic labels, accuracy metrics are not directly comparable. We therefore focus on representation cost and end-to-end latency on Raspberry Pi~5 as reported/measured under each study's respective setup. This is not a controlled head-to-head benchmark; the comparison is intended to contextualize the runtime implications of occupancy-grid construction versus compact scan encoding.

\begin{table*}[!t]
\caption{Comparison to a Close 2D LiDAR--YOLO Occupancy-Grid Pipeline}
\begin{center}
\begin{tabular}{lcccccc}
\hline
\textbf{Method} & \textbf{Representation} & \textbf{Input} & \textbf{Data regime} & \textbf{RPi~5 time (ms, end-to-end)} & \textbf{Derived FPS} & \textbf{mAP@0.5} \\
\hline
Najem et al.~\cite{najem2024object} & Occupancy grid & $640\times640$ & Real + manual labels & 846.5 & 1.18 & --- \\
Our method & 3-frame RGB tensor & $64\times384\times3$ & Sim + auto labels & 47.8 & 20.92 & 0.984 \\
\hline
\end{tabular}
\label{tab:comparison}
\end{center}
\noindent \footnotesize Derived FPS values are computed as \(1/\text{reported time}\).
\end{table*}

As shown in Table~\ref{tab:comparison}, the occupancy-grid pipeline in~\cite{najem2024object} reports substantially higher end-to-end latency on Raspberry Pi~5 than our compact scan encoding. This should be interpreted cautiously because the datasets, labeling regimes, software stacks, and timing protocols differ; nevertheless, the direction of the difference is consistent with the added preprocessing cost of occupancy-grid construction and the larger network input size.

\section{Conclusion and Future Work}

This paper presented a privacy-friendly, real-time 2D LiDAR object detection pipeline for indoor service robots. The method encodes short-term temporal context by rasterizing three consecutive scans and stacking them as RGB channels, producing a compact $64\times384\times3$ input compatible with YOLOv8n while avoiding occupancy-grid construction, which can add preprocessing overhead. In Webots with strict scenario-level holdout, the proposed approach achieves 98.4\% mAP@0.5 (0.778 mAP@0.5:0.95) with 94.9\% precision and 94.7\% recall on four indoor object classes. On Raspberry Pi~5, the full pipeline runs in real time with a mean post-warm-up end-to-end latency of 47.8~ms/frame (including encoding and postprocessing), indicating that lightweight temporal encoding can support accurate LiDAR-only detection on embedded hardware.

Future work will evaluate sim-to-real transfer on physical robots with real 2D LiDAR sensors, including robustness to noise, missing returns, and calibration effects. We also plan to extend the object set (including dynamic agents), study failure modes under fast rotations where unaligned temporal stacking may degrade, and explore deployment optimizations such as quantization and hardware-accelerated runtimes to further reduce latency and power consumption.

\FloatBarrier
\bibliographystyle{IEEEtran}
\bibliography{bib}

\end{document}